\documentclass{moriond}

\def\be{\begin{equation}}
\def\ee{\end{equation}}
\def\bea{\begin{eqnarray}}
\def\eea{\end{eqnarray}}

\newcommand{\Photo}{\includegraphics[height=35mm]{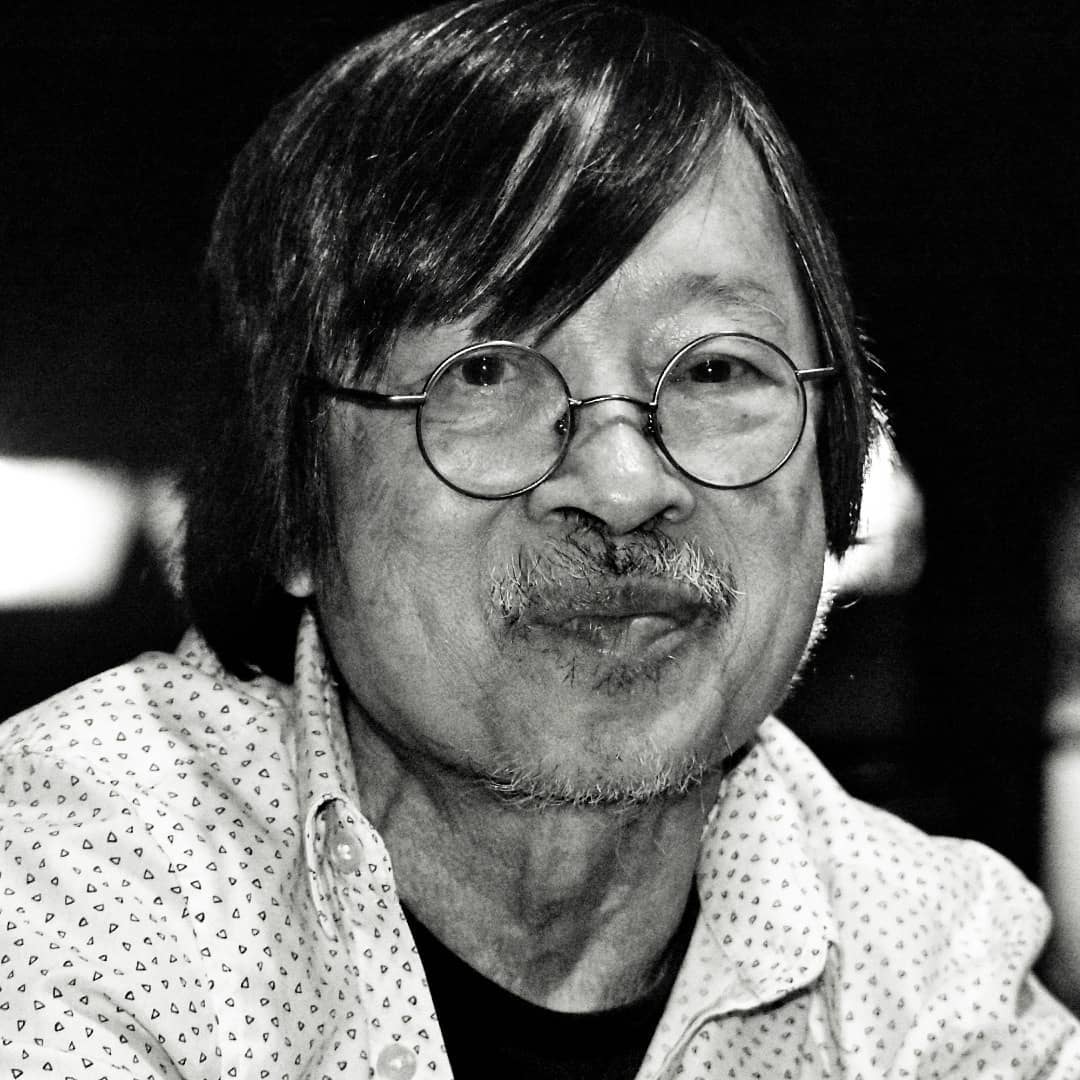}}

\begin{document}
\vspace*{4cm}
\title{THE LIFETIME FRONTIER: SEARCH FOR NEW PHYSICS WITH LONG-LIVED PARTICLES}

\author{ P.Q. HUNG }

\address{Department of Physics, University of Virginia,
Charlottesville, VA 22904-4714, USA}

\maketitle\abstracts{
The search for new physics with long-lived particles is an ongoing and thriving effort in the High Energy Physics community which necessitates new search strategies such as novel algorithms, novel detectors, etc...For these reasons, one could perhaps add another frontier, the Lifetime Frontier, to the standard three (Energy, Intensity and Cosmic). In this talk, I will describe a BSM physics model whose characteristic signatures are decays of new (mirror) fermions at displaced vertices, a domain belonging to the Lifetime Frontier. It is a model of {\em non-sterile} right-handed neutrinos whose masses are proportional to the electroweak scale $\Lambda_{EW} \sim 246 \, GeV$. The model proposed a solution to the strong CP problem with a surprising connection between the sizes of the neutrino masses and the $\theta$-angle which contributes to the neutron electric dipole moment.} 

\section{Introduction}
The discovery of the 125-GeV scalar and the absence of any sign of new physics at the LHC have led the high energy physics community to ponder the question whether or not there is any new physics at all or whether one should build a bigger accelerator and hope to find it there. It goes without saying that the SM as it now stands is incomplete. How incomplete is it? Leaving aside dark matter about which we have very little information apart from its gravitational effects on galactic scales, there already exists quantities that we have already measured experimentally but whose origins are still unknown: the hierarchy of quark and lepton masses and mixings, the nature of neutrino masses which is inferred from neutrino oscillation data, and even the 125-GeV scalar itself (are there more scalars?).

Most searches for new physics concentrated on regions of the detector which are close to the collision point in the beam pipe at distances $< O(1mm)$ for prompt particle decays and regions which are on the edge of the detector for stable particorles. The present absence of any sign of new physics might prompt us to ask whether or not we are not looking at the right region. Perhaps it (new physics) is hidden somewhere in regions of the detector which are not included in the present search algorithms, e.g. at distances greater than a few millimeters? If we are looking for new particles through their decays, this unexplored region would correspond to typical places where long-lived particles (LLP) would decay. We shall call this region "The X-Files" after a very popular TV series. This is the story of FBI investigations of unusual, paranormal phenomena which are classified in the so-called X-Files. The two FBI agents involved in the investigations are Fox Mulder and Dana Scully. Fox Mulder believes in the existence of extraterrestrials who are involved in these paranormal phenomena. Dana Scully tries to debunk Mulder's speculations. The parallelism with High Energy Physics is very suggestive: Fox Mulder $\leftrightarrow$ Theorists; Paranormal phenomena $\leftrightarrow$ Displaced vertices,...; Extraterrestrials $\leftrightarrow$ New Physics; Dana Scully $\leftrightarrow$ Experimentalists. In the "X-Files" are speculative models such as L-R Models, R-parity violating SUSY, Split SUSY,..., Origins of \underline{neutrino masses}. 

Among the various items in the aforementioned X-Files, "neutrino masses" is underlined. Although its origin is still unknown, the fact that neutrinos have a mass is indisputable and this is the only evidence we have so far of physics beyond the SM (BSM). But why would it be classified in the X-Files, the region of LLP? What aspects of neutrinos are we searching for in that region? To answer this question, let us ask another question: How do {\em tiny} neutrino masses come about?

There are many proposals for the existence of {\em tiny} neutrino masses with the most attractive and popular one being the so-called {\em seesaw mechanism}. Because of space limitation, I will not be able to do justice to all proposals on the market but will concentrate instead on the seesaw mechanism which basically postulates the existence of right-handed neutrinos $\nu_R$.

Basically the seesaw mechanism involves {\em two} mass terms: the Dirac mass term $m_D \bar{\nu}_L \nu_R +H.c.$, and the Majorana mass term $M_R \nu_R^T \nu_R$. It is worthwhile to repeat a well-known fact: the aforementioned mass terms involve a mixing between left-handed and right-handed neutrinos. If one assumes $m_D (Dirac) \ll M_R (Majorana)$, the diagonalization of a 2x2 matrix yields the iconic eigenvalues: $|m_\nu| =m_D^2/M_R  \sim O(<eV)$ and $M_R=?$. What are the physics mechanisms behind the generation of $m_D$ and $M_R$? How massive are $\nu_R$'s and do they interact with the W and Z bosons of the SM?

In standard seesaw scenarios, $\nu_R$'s are SM singlets and are thus {\em sterile}. In grand unified models where the SM is embedded  such as SO(10) for example, $M_R$ is naturally of O($\Lambda_{GUT} \sim 10^16 \, GeV$) and $\nu_R$'s are thus inaccessible both from the point of view of production- they do not couple to SM gauge bosons- and from the point of view of mass scale. More involved models can lower $M_R$ significantly but not enough for $\nu_R$'s to be produced at earth-bound accelerators. Another class of models, the left-right symmetric model $SU(3)_C \times SU(2)_L \times SU(2)_R \times U(1)_{B-L}$ \cite{LR},  where $\nu_R$'s are $SU(2)_L$-singlets but are members of $SU(2)_R$-doublets, has more potential for the discovery of $\nu_R$'s. However, this potential is intrinsically linked to the discovery of $W_R$, the gauge bosons of $SU(2)_R$. Notice that, in all of these models, $m_D$ is proportional to $\Lambda_{EW} \sim 246 \, GeV$.
It goes without saying that this brief description of $\nu_R$'s  as sterile particles is very far from doing justice to works on the subject.

WHAT IF $\nu_R$'s are {\em non-sterile} and interact with W and Z with the same strength as that of other known SM quarks and leptons? WHAT IF $M_R$ is proportional to $\Lambda_{EW}$? Two remarks are in order at this point. First, there is NO physical principle forbidding a non-sterile $\nu_R$'. Second, this could be a TESTABLE  scenario. This is the essence to the electroweak-scale non-sterile right-handed neutrino model or simply the EW-$\nu_R$ model \cite{pqnur}.

\section{The EW-$\nu_R$ modelwith LLPs}

Below is a brief summary of the gauge and particle contents of the model  \cite{pqnur}, as well as interactions relevant for the discussion presented in this talk.

1) Gauge group: The same gauge group as in the SM i.e. $SU(3)_C \times SU(2)_W \times U(1)_Y$.

2) Fermions: SM: $l_L = \left(
	  \begin{array}{c}
	   \nu_L \\
	   e_L \\
	  \end{array}
	 \right)$; $q_L = \left(
	  	 \begin{array}{c}
	   	  u_L \\
	     	  d_L \\
	  	\end{array}
	 	\right)$; $e_R; \ u_R, \ d_R$;

Mirror: $l_R^M = \left(
	  \begin{array}{c}
	   \nu_R^M \\
	   e_R^M \\
	  \end{array}
	 \right)$; $q_R^M = \left(
	  	 \begin{array}{c}
	   	  u_R^M \\
	     	  d_R^M \\
	  	\end{array}
	 	\right)$; $e_L^M; \ u_L^M, \ d_L^M$.
		
3) Scalars: 

\underline{Doublet Higgs fields} \cite{pqnur,pqcpnew} (similar to 2HDM): $\Phi^{SM}_{1} (Y/2=-1/2)$, $\Phi^{SM}_{2} (Y/2=+1/2)$ coupled to SM fermions and $\Phi^{M}_{1} (Y/2=-1/2)$, $\Phi^{M}_{1} (Y/2=+1/2)$ coupled to mirror fermions with $\langle \Phi^{SM}_{1} \rangle=(v_1/\sqrt{2},0)$,  $\langle \Phi^{SM}_{2} \rangle=(0,v_2/\sqrt{2})$ and $\langle \Phi^{M}_{1} \rangle=(v^M_1/\sqrt{2},0)$, $\langle \Phi^{M}_{2} \rangle=(0,v^M_2/\sqrt{2})$ .

\underline{Triplet Higgs fields}: One complex $\tilde{\chi}=(\chi^0,\chi^+,\chi^{++})$ and one real $\xi (Y/2 = 0)= (\xi^+, \xi^0, \xi^-)$ which can be put in a 3x3 matrix of global $SU(2)_L \times SU(2)_R$

$\chi = \left( \begin{array}{ccc}
\chi^{0} &\xi^{+}& \chi^{++} \\
\chi^{-} &\xi^{0}&\chi^{+} \\
\chi^{--}&\xi^{-}& \chi^{0*}
\end{array} \right)$

with $\langle \chi^0 \rangle = \langle \xi^0 \rangle=v_M$ in order to preserve Custodial Symmetry (that guarantees $M_{W}^2 = M_{Z}^2 \cos^{2}\theta_W$ at tree level). Here $(\sum_{i=1,2} v_i^2 + v^{M,2}_{i}) + 8 v_M^2 = (246 GeV)^2$.	 

\underline{Singlet Higgs fields}:
$\phi_S$ with $\langle \phi_S \rangle =v_S \ll v_M$ (to be explained below): Important scalars connecting the SM and Mirror worlds. Crucial in the search for mirror fermions $\rightarrow$ displaced vertices. Crucial for the strong CP problem.

4) Relevant interaction Lagrangians:
\begin{itemize}
\item 
\be
\label{Majorana}
L_M =  g_M \, l^{M,T}_R  \sigma_2 \tau_2 \ \tilde{\chi} \ l^M_R,
\ee
giving $M_R = g_M v_M $ $\Rightarrow$ $M_Z/2<M_R < O(\Lambda_{EW} \sim 246 GeV)$. Energetically accessible!
\item 
\be
\label{Dirac}
{\mathcal L}_S = - g_{Sl} \,\bar{l}_L \ \phi_S \ l_R^M + {\rm H.c.},
\ee
giving $m_D = g_{Sl} \ v_S $. This is an interaction that is crucial in the decay of a mirror lepton into a SM lepton plus a singlet scalar field. Obviously accessible but where is it? It depends on $g_{Sl}$! . For the quark sector, $g_{Sl} \rightarrow g_{Sq}$  (from $g_{Sq} \bar{q}^{M} \phi_{S} q_L$).

One now has $|m_\nu| =m_D^2/M_R  = (g_{Sl}^2/g_M)(v_S^2/v_M) \sim O(<eV)$. What does all this have to do with the X-Files scenarios? As can be seen above, mirror fermions decaying into a SM fermion and a singlet scalar will have a lifetime that depends on the Yukawa couplings $g_{Sl}$ and $g_{Sq}$. It turns out, as we will show next, that these Yukawa couplings are restricted experimentally to be very small. As a result, the decays of the lightest mirror fermions into SM fermions will always occur at {\em displaced vertices}. In consequence, the {\em lightest mirror fermions} are the LLPs in the EW-$\nu_R$ model .

\end{itemize}

\section{Experimental constraints on $g_{Sl}$}

The contributions to the processes $\mu \rightarrow e \gamma$ \cite{muegamma}, $\mu$ to $e$ conversion \cite{mutoe} come from the interaction of Eq.~\ref{Dirac} where the muon is changed into a mirror lepton and a singlet scalar $\phi_S$ in the one-loop diagrams. 
\begin{figure}
\centering
\centerline{
\includegraphics[width=0.7\linewidth]{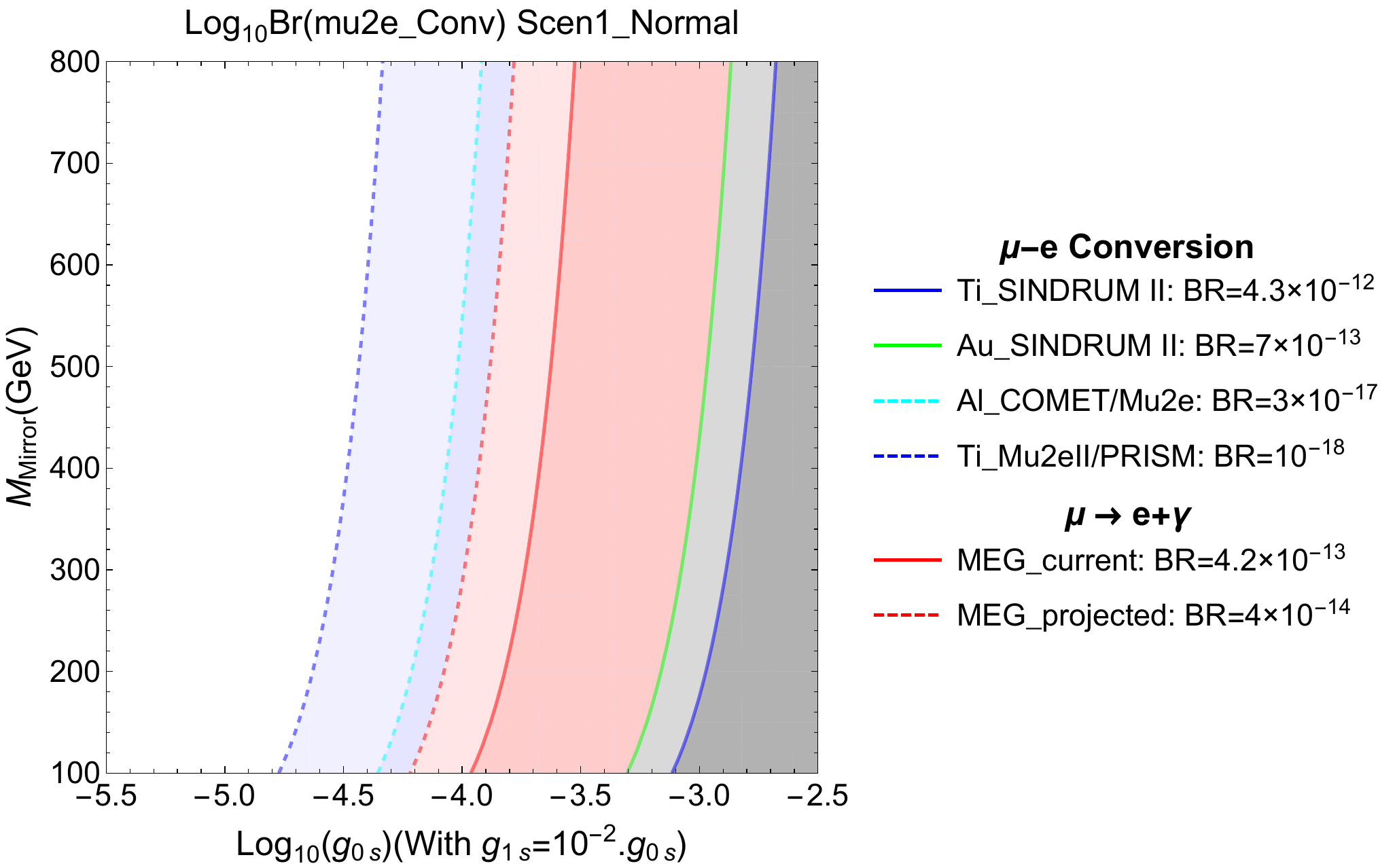}} 
\caption{{\small Experimental constraints from $\mu \rightarrow e \gamma$, $\mu$ to $e$ conversion }}
\label{fig:rare}
\end{figure}
The results are summarized in Fig.~\ref{fig:rare} \cite{muegamma,mutoe}.
From Fig.~\ref{fig:rare}, the strongest constraint comes from the MEG $\mu \rightarrow e \gamma$ experiment. Typically, $g_{Sl}< 10^{-4}$.

\section{Experimental constraints on $g_{Sq}$}

The constraint on $g_{Sq}$ comes indirectly from a proposed axionless solution to the strong CP problem. A very brief summary of that problem is in order here.

The QCD vacuum is complicated as expounded by 't Hooft \cite{thooft}and the proper way to write a gauge-invariant vacuum is to characterize it by an "angle" $\theta$, namely $| \theta \rangle = \sum_{n} \exp (-\imath n \, \theta) |n \rangle$. This has the effect of changing the action as follows $S_{eff} = S_{gauge} + \theta \, (g_{3}^2/32 \pi^2) \int d^ x \, G_{a}^{\mu \nu} \tilde{G_{\mu \nu}^{a}}$. The integrand of the second term in the aforementioned equation behaves like $\vec{E}.\vec{B}$ where $\vec{E}$ and $\vec{B}$ have opposite signs under a CP transformation. As a result, it is CP-violating. This term contributes to the electric dipole. moment of the neutron as $d_n \approx 5.2 \times 10^{-16} \theta_{QCD} \,  e-cm$. Experimentally, $|d_n| < 2.9 \times 10^{-26} \, e-cm$ leading to the constraint $\theta < 10^{-10}$. Why is it so small? That is the strong CP problem. It is made worse by the fact that
in general quarks mass matrices are {\em complex} and introduces an extra "angle" $Arg Det M$, i.e. the effective angle appearing in front of the CP-violating integral is now

\be
\label{theta}
\bar{\theta} = \theta + Arg Det M
\ee

The most famous solution to the strong CP problem is without question the Peccei-Quinn axion \cite{P-Q}. By imposing an extra chiral global symmetry $U(1)_{PQ}$, Peccei and Quinn argued that a chiral rotation by a phase $\sigma$ changes the $\theta$-vacuum to another vacuum characterized by $\theta-2\sigma$ with the same energy: one has an equivalent theory for any value of $\theta$ meaning that it is "irrelevant" \cite{jackiw}. One can "rotate" away the original angle to zero and has a CP-conserving theory provided  $Arg Det M=0$. This happens in the Peccei-Quinn model when the scalar has vanishing VEV. In realistic theories, one expects $Arg Det M \neq 0$. The Peccei-Quinn solution is to replace it by a dynamical field $Arg Det M \rightarrow a(x)/f$ which settles to zero at the minimum of the associated potential. Forty years after its conception, the axion is still the subject of intense experimental (earth-bound and astrophysical) searches.

There are several axionless models for the strong CP problem. It turns out that the EW-$\nu_R$ model has a built-in global symmetry  $U(1)_{SM} \times U(1)_{MF}$ which is imposed to prevent unwanted terms which could disturb the seesaw mechanism. This global symmetry has a chiral part which can nicely be used to "rotate" away $\theta$ leaving behind the extra contribution from the quark mass matrix $Arg Det M$. A calculation within a toy model of one family of quarks and one family of mirror quarks which can be generalized to the more realistic case of three families shows the following result \cite{pqcpnew,pqcpold}. With $\bar{\theta} \equiv \theta_{Weak} = Arg Det ({\cal M}_u {\cal M}_d)$ 
\be
\label{thetaw}
 \theta_{Weak}  \approx  -(r_u \sin(\theta_q + \theta_u) + r_d \sin(\theta_q + \theta_d)) ,
\ee
where $r_u=(|g_{Sq}||g_{Su}|/g_{Sl}^2)(m_D^2/(m_u M_u))$; $r_d= |g_{Sq}||g_{Sd}|/g_{Sl}^2)(m_D^2/(m_d M_d))$ with $m$ and $M$ denoting the SM and mirror quark masses respectively. As seen from the aforementioned results, the non-vanishing $\bar{\theta}$ arises from the mixing between SM and mirror quarks through the singlet scalar $\phi_S$ and the phases reside in the off-diagonal matrix elements of the mass matrices. As noticed in \cite{pq_CP}, assuming the Yukawa couplings $g_{Sq,u,d,l} \neq 0$, $m_D \rightarrow 0$ as the VEV of $\phi_S$, namely $v_S$, goes to {\em zero}, i.e. $ \theta_{Weak}  \rightarrow 0$ (just like the P-Q scenario) when $m_{\nu} \rightarrow 0$. Since $m_{\nu} \sim O(eV) \neq 0$, $\bar{\theta}$ is small because $m_{\nu}$ is small! It {\em does not} have to be zero! This has interesting implications on future measurements of the neutron electric dipole moment. Putting reasonable numbers in Eq.~\ref{thetaw}, one gets $\theta_{Weak} < -10^{-8}\{(\frac{|g_{Sq}||g_{Su}|}{g_{Sl}^2})\sin(\theta_q + \theta_u)  + (\frac{|g_{Sq}||g_{Sd}|}{g_{Sl}^2})\sin(\theta_q + \theta_d)\}$. Without fine-tuning, this implies that 
\be
\label{gsq}
|g_{Sq}| < |g_{Sl}| < 10^{-4} \, .
\ee
Mirror quarks are long-lived too!

\section{Search for long-lived mirror fermions}

\subsection{Long-lived mirror leptons}

The signals to look for are lepton-number violating signals at high energy: Like-sign dileptons from the decays $\nu_R \nu_R$ ($q\bar{q} \rightarrow Z \rightarrow \nu_R \nu_R)$. Remember that $\nu_R$s are {\em non-sterile and Majorana}! One has
$\nu_{Ri} \rightarrow e_{Rj}^{M} + W^+$ followed by $e_{Rj}^{M} \rightarrow e_{Lk} + \phi_S$ which occurs at {\em displaced vertices} due to the smallness of $g_{Sl} < 10^{-4}$.
The signals at the LHC would be $q\bar{q} \rightarrow Z \rightarrow {\nu_Ri} + \nu_{Ri} \rightarrow e_{Lk}+ e_{Ll}+ W^+ + W^+ +\phi_S + \phi_S$: Like-sign dileptons  $e_{Lk}+ e_{Ll}$ plus 2 jets (from 2 $W$) plus missing energies (from $\phi_S$) $\Rightarrow$ Lepton-number violating signals! The appearance of like-sign dileptons ($e^{-}e^{-},\mu^{-}\mu^{-}, \tau^{-} \tau^{-}, e^{-} \mu^{-},...$) could occur at  {\em displaced vertices} $> 1mm$ or even tens of centimeters depending on the size of $g_{Sl}$. One can also have $\bar{u}+d \rightarrow W^{-} \rightarrow e_{Rj}^{M} + \nu_R \rightarrow e_{L}+ e_{L}+1 Jet+\phi_S + \phi_S$. See also the discussion in \cite{chakdar1}.	


\subsection{Long-lived mirror quarks}

A typical decay chain of a mirror quark starts with the heaviest one which decays into a lighter one plus a W, e.g. $u^{M}_i \rightarrow d^{M}_j +W$. The lightest mirrow quark can only decay into a SM quark plus the singlet scalar $\phi_S$.  Since the decay length of a "free" lightest mirror quark ($> 1mm$) is typically {\em much larger} than a hadronic length ($\sim 1 fermi$), the gluon fusion process first gives rise to a mirror meson ($\bar{q}^M q^M$) which subsequently decays, at a {\em displaced vertex}, into a pair of SM quark and antiquark. Since mirror quarks are heavy, the formation of a mirror meson can be accomplished, to a good approximation, by a QCD Coulomb-like potential $V(r) \approx -4\alpha_{s}(r)/3r$. It is straightforward to compute the mirror meson wave function at the origin which will allow us to obtain the production cross section and the decay rate (details can be found in \cite{dat}). Basically, the relevant process is $\sigma (gg \rightarrow \eta^M \rightarrow q \bar{q})$ where $\eta^M$ stands for "mirror meson". In $\eta^M \rightarrow q \bar{q}$, there is {\em no missing energy} in the form of $\phi_S$ because $q^M$ and $\bar{q}^M$ exchange $\phi_S$ and transform into $q$ and $\bar{q}$. (For an earlier discussion, see \cite{chakdar2}.)

From Fig.~2, one notices that mirror mesons decay {\em well inside} a typical silicon vertex detector  and beyond for a range of mass ($m_{\eta^M} \approx 2 m_{q^M}$) and for $g_{Sq}< 10^{-4}$. The lightest mirror mesons are long-lived! Other kinds of mesons and even "baryons" involving one or more mirror quarks are under study.
\begin{figure}
\label{figure2}
\centering
    \includegraphics[scale=0.48]{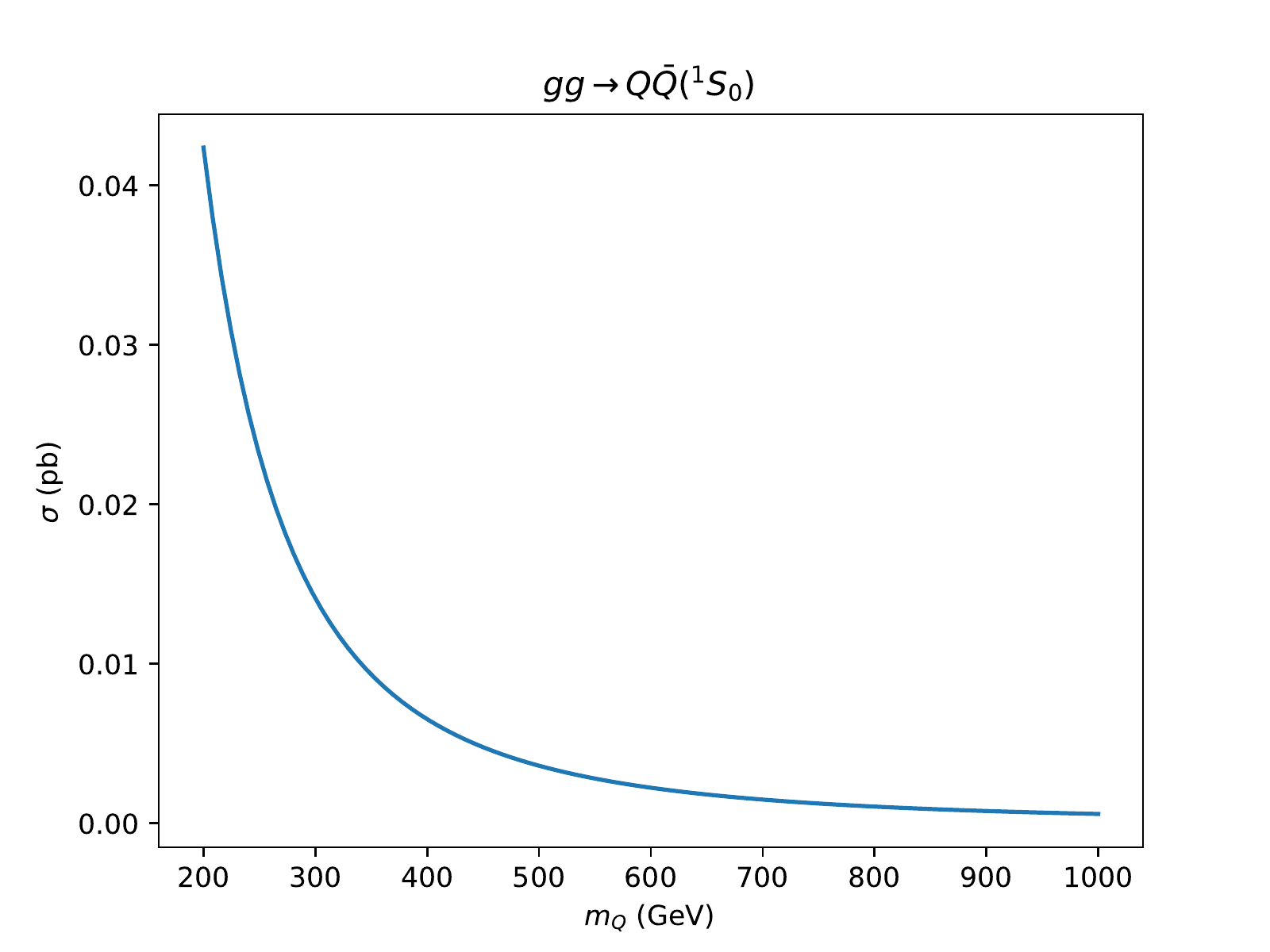} 
    \includegraphics[scale=0.48]{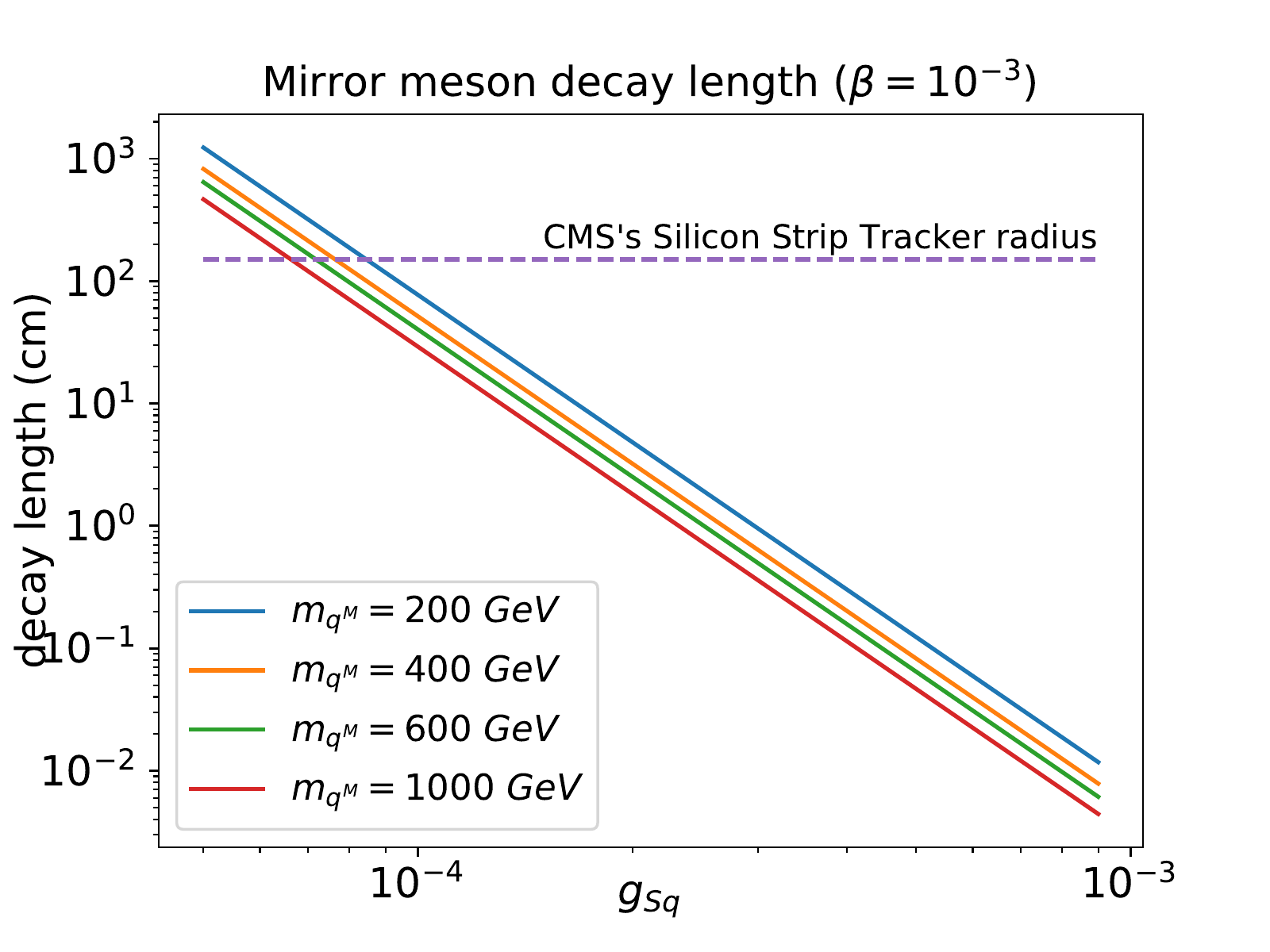} 
\caption{{\small Production cross section and decay length of mirror mesons }}
\end{figure}
\section{Conclusions}
\begin{itemize}
\item The EW-$\nu_R$ model belongs to a class of models where characteristic signatures are {\em Long-lived particles}. They are the lightest mirror fermions.

\item The case of whether neutrinos are of a Dirac or Majorana nature can be settled if like-sign dilepton signals are {\em discovered at displaced vertices} at hadron colliders such as the LHC. Is it a harder or "easier" experiment as compared with $0\nu \beta \beta$ experiments? In some sense, like-sign dilepton "signals" at displaced vertices are the high-energy equivalence of low-energy $0\nu \beta \beta$ "signals".

\item There seems to be a deep connection between neutrino physics and QCD in the solution to the strong CP problem: The magnitude of $\bar{\theta}$ is linked to the magnitude of the neutrino mass, namely {\em very small}.  

\item The Lifetime Frontier promises interesting windows into physics beyond the SM.

\end{itemize}

\end{document}